\documentclass[12pt,draftcls,onecolumn]{IEEEtran}
\usepackage{graphicx}  % Written by David Carlisle and Sebastian Rahtz
\begin{document}

\title{Theoretical Analysis of Tuned HVAC Line for Low Loss Long Distance Bulk Power Transmission
\vspace{0.5in}}
\author{{Abhisek Ukil}\\
{(corresponding author)\\
School of Electrical and Electronic Engineering\\
Nanyang Technological University (NTU)\\
639798, Singapore\\
Phone: +65 6790 5374\\
Fax: +65  6793 3318\\
Email: aukil@ntu.edu.sg}}

\maketitle

\newpage
\begin{abstract}
One of the main objectives of the smart grid initiative is to enable bulk power transmission over long distance, with reduced transmission losses. Besides the traditional high-voltage alternating current (HVAC) transmission, with the advancement in power electronics, high-voltage direct current (HVDC) transmission is increasingly becoming important. One of the main factors impacting the transmission line parameters and the losses is the length of the transmission line (overhead). In this paper, a concept of tuned high-voltage AC line is presented for long ($>250$ km) transmission line. A tuned line is where the receiving-end voltage and current are numerically equal to the corresponding sending-end values. This paper presents the detailed theoretical analysis of the tuned HVAC line, suggesting adaptation of the transmission frequency as per the length of the line. The simulation of a tuned HVAC line is performed using the PSCAD/EMTDC. Simulation results for two different line lengths, substantiate the theoretical analysis of reducing the reactive power absorbed in the line, while increasing the active power transmission.

\end{abstract}



% for peerreview papers, inserts a page break and creates the second title.
% Will be ignored for other modes.
%\IEEEpeerreviewmaketitle

\newpage

\section{Introduction}
Transmission of bulk electric power over long distance at reasonable loss plays the pivotal role in power transmission and distribution. This has always been an important research arena in the power systems domain. The length of the transmission line (overhead) plays an important role in the power transmission, impacting the line parameters, and the associated losses. With the advent of modern technologies, namely due to advancement in power electronics, besides the traditional high-voltage alternating current (HVAC) transmission, high-voltage direct current (HVDC) transmission is becoming increasingly important.  

An HVDC electric power transmission system uses direct current for the transmission of electrical power. The modern form of HVDC transmission uses technology developed extensively in the 1930s in Sweden (ASEA) and in Germany. Early commercial installations included one in the Soviet Union in 1951 between Moscow and Kashira and between the island of Gotland and Swedish mainland in 1954 \cite{Wang13}. The previous longest HVDC link in the world was the Xiangjiaba-Shanghai 2,071 km, $\pm$800 kV, 6400 MW link connecting the Xiangjiaba Dam to Shanghai, in the People's Republic of China. Early in 2013, the longest HVDC link has been the Rio Madeira link in Brazil, which consists of two bipoles of $\pm$600 kV, 3150 MW each, connecting Porto Velho in the state of Rondonia to the Sao Paulo area, over a length of 2,375 km. Rock Island Clean Line was being installed in North America, over a length of 805 km, and power of 3,500 MW which is expected to be completed in the year 2017 \cite{Wang13}. 

There has been increasing focus on the implementation of multi-terminal HVDC system (69-765 kV range) \cite{Jovcic10}--\cite{Bahrman07}, as it is deemed to facilitate efficient integration of the distributed energy resources (DERs) into the grid. With the advanced power electronic converter, e.g., voltage-source converters (VSCs) \cite{Tang14}--\cite{HVDC2}, HVDC technology is becoming more feasible and promising. Modern day HVDC systems use either the traditional thyristor-based current source converter (CSC) technology, or self-commutated voltage source converter (VSC) technology. VSC is based on insulated-gate bipolar transistors (IGBTs). Comparison of the CSC and the VSC-based HVDC systems can be referred to in \cite{HVDC1},\cite{HVDC2},\cite{Franck11}. Protection is still a major issue in the HVDC system, especially for multi-terminal HVDC system. This includes relatively immature DC circuit-breaker technology \cite{Franck11}--\cite{Alstom_HVDC}, as well as lack of robust protection logic, e.g., for differentiating the fault and the load changes in the HVDC system \cite{Ukil1}--\cite{Ukil3}.

In this paper, a concept of tuned HVAC line is presented for long transmission line. The term `tuned' is linked with the length of line to the transmission frequency, explained latter in the paper. The simulation results show that the tuned HVAC line could effectively reduce the reactive power absorbed in the line. For long distance power transmission, the tuned HVAC system could provide an alternative solution to the HVDC system.

The remainder of the paper is organized as follows. Section II presents the detailed theoretical analysis, consisting of analysis of long transmission line, tuned HVAC line, and reactive power. Simulation results for the tuned HVAC line are presented in section III, followed by conclusions in section IV. 

\section{Theoretical Analysis}
\subsection{Review of Analysis of Long Transmission Line}
\begin{figure}[ht]
\centering
\includegraphics[width=4.5in]{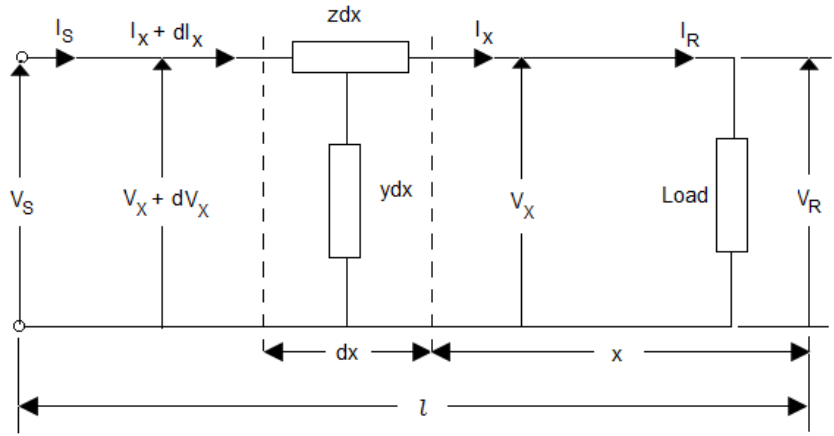}
\caption{Schematic diagram of long ($>250$ km) transmission line.}
\label{fig:Longline}
\end{figure}

For short ($<80$ km) and medium ($<250$ km) length transmission lines, the transmission line parameters, e.g., the series impedance and the shunt admittance can be modeled as lumped parameter. However, for the long ($>250$ km) lines, those parameters cannot be assumed to be lumped, rather distributed uniformly throughout the length of the transmission line \cite{Grainger94},\cite{Nagrath94}. Fig. \ref{fig:Longline} shows the schematic diagram of the long transmission line.

In Fig. \ref{fig:Longline}, $V_{S}$ and $V_{R}$ represent the sending- and the receiving-end voltages respectively, $I_{S}$ and $I_{R}$ the sending- and the receiving-end current flows. For the elemental line section $dx$ of the total length $l$, the series impedance is represented as $zdx$, and the shunt admittance as $ydx$. Then, using the general formulation of the transmission line \cite{Grainger94},\cite{Nagrath94}, the following relationship holds between the voltage and the current of the sending- and the receiving-ends.

\begin{equation}
\label{eq:Longline}
\left[\begin{array}{c}
V_{S} \\
I_{S}
\end{array}\right]=
\left[\begin{array}{c}
\cosh(\gamma l) \quad Z_{c}\sinh(\gamma l) \\
\frac{1}{Z_{c}}\sinh(\gamma l) \quad \cosh(\gamma l)
\end{array}\right]
\left[\begin{array}{c}
V_{R} \\
I_{R}
\end{array}\right]
\end{equation}

where, $Z_{c}$ is the characteristic impedance of the line, and $\gamma$ is the propagation constant \cite{Grainger94},\cite{Nagrath94}. For overhead lines, the shunt conductance and the line resistance can be considered negligible. Therefore, only the line inductance ($L$) and the capacitance ($C$) would count.

\begin{equation}
\label{eq:gamma}
\gamma = \sqrt{yz} = j\omega \sqrt{LC},
\end{equation}

where, $\omega$ is the line frequency (for AC). Now, the following expressions hold,
\begin{equation}
\label{eq:cosh}
\cosh(\gamma l) =\cosh( j\omega l \sqrt{LC})= \cos (\omega l \sqrt{LC}),
\end{equation}

\begin{equation}
\label{eq:sinh}
\sinh(\gamma l) =\sinh( j\omega l \sqrt{LC})= j\sin (\omega l \sqrt{LC}).
\end{equation}

Therefore, the eq. (\ref{eq:Longline}) simplifies to
\begin{equation}
\label{eq:LonglineSimple}
\left[\begin{array}{c}
V_{S} \\
I_{S}
\end{array}\right]=
\left[\begin{array}{c}
\cos(\omega l \sqrt{LC}) \quad jZ_{c}\sin(\omega l \sqrt{LC}) \\
\frac{j}{Z_{c}}\sin(\omega l \sqrt{LC}) \quad \cos(\omega l \sqrt{LC})
\end{array}\right]
\left[\begin{array}{c}
V_{R} \\
I_{R}
\end{array}\right].
\end{equation}

\subsection{Analysis of Tuned HVAC Line}
A tuned line is where the receiving-end voltage and current are numerically equal to the corresponding sending-end values \cite{Grainger94},\cite{Nagrath94}. That is,

\begin{equation}
\label{eq:tuned_V}
|V_{S}|=|V_{R}|,
\end{equation}

\begin{equation}
\label{eq:tuned_I}
|I_{S}|=|I_{R}|.
\end{equation}

From the simplified general relation in eq. (\ref{eq:LonglineSimple}),

\begin{equation}
\label{eq:voltages}
V_{S}=V_{R}\cos(\omega l \sqrt{LC}) + jI_{R}Z_{c}\sin(\omega l \sqrt{LC}).
\end{equation}

Using eq. (\ref{eq:voltages}), the condition for the tuned line in eq. (\ref{eq:tuned_V}) can be possible if

\begin{equation}
\label{eq:currentzero}
|I_{R}|Z_{c}\sin(\omega l \sqrt{LC})=0.
\end{equation}

Now, for a valid receiving-end current ($|I_{R}|\neq 0$), and a line with non-zero characteristic impedance, eq. (\ref{eq:currentzero}) holds only if,
\begin{equation}
\label{eq:sinzero}
\sin(\omega l \sqrt{LC})=0, \Rightarrow \omega l \sqrt{LC}=n\pi,  n=1,2,3,\ldots .
\end{equation}

For an AC line with frequency $f$ Hz, from eq. (\ref{eq:sinzero}), the length of the line for tuning is

\begin{equation}
\label{eq:linelength}
l=\frac{n}{2f\sqrt{LC}}.
\end{equation}

Since, $\frac{1}{\sqrt{LC}}$ is almost equal to the velocity of light $c$ ($=3\times 10^{5}$ km/s), for power frequency, e.g., 50 Hz, the tuned line lengths can be calculated using the eq. (\ref{eq:linelength}) as 3000 km, 6000 km, 9000 km, $\ldots$ (corresponding to $n=1,2,3,\ldots$). The minimum length of 3000 km line is considered to be uneconomical due to the transmission losses, e.g., due to the absorbed reactive power, etc.

Now, eq. (\ref{eq:linelength}) can be reordered as
\begin{equation}
\label{eq:linelengthfreq}
f=\frac{nc}{2l}.
\end{equation}

Therefore, for a long line, e.g., $l=500$ km, the tuning frequency would be 300 Hz, 600 Hz, 900 Hz, $\ldots$ (corresponding to $n=1,2,3,\ldots$). Thus, for a given line length, eq. (\ref{eq:linelengthfreq}) can be used for calculating the optimal transmission frequency for a tuned, i.e., low loss (ideally loss-less) condition.

From eq. (\ref{eq:tuned_V}), it can be noticed that for the tuned HVAC line, the voltage regulation will theoretically be zero.

\subsection{Analysis of Reactive Power}
The active power ($P_{R}$) and the reactive power ($Q_{R}$) at the receiving side (see Fig. \ref{fig:Longline}) are given by the following relations, \cite{Grainger94},\cite{Nagrath94}.

\begin{equation}
\label{eq:activepower}
P_{R}=\frac{\left|V_{S}\right| \left|V_{R}\right| \sin\delta}{X},
\end{equation}

\begin{equation}
\label{eq:reactivepower}
Q_{R}=\frac{\left|V_{S}\right| \left|V_{R}\right| \cos\delta}{X} - \frac{\left|V_{R}^{2}\right|}{X},
\end{equation}
where, $\delta$ is the torque angle and $X$ is the line reactance. 

We define the voltage regulation as
\begin{equation}
\label{eq:voltreg}
\Delta V=\frac{\left|V_{S}\right|-\left|V_{R}\right|}{\left|V_{R}\right|}.
\end{equation}

Then the reactive power in eq. (\ref{eq:reactivepower}) becomes

\begin{equation}
\label{eq:reactivepower_voltreg}
Q_{R}=\frac{\left|V_{R}^{2}\right|\left[\left(1+\Delta V\right)\cos\delta - 1\right]}{X}.
\end{equation}

For the tuned HVAC line, as $|V_{S}|=|V_{R}|$, and $\Delta V=0$ (see eq. (\ref{eq:tuned_V})), the reactive power in eq. (\ref{eq:reactivepower_voltreg}) becomes,

\begin{equation}
\label{eq:reactivepower2}
Q_{R,TUNED}=\frac{\left|V_{R}^{2}\right|}{X}(\cos\delta - 1).
\end{equation}

Comparing eqs. (\ref{eq:reactivepower_voltreg} \& \ref{eq:reactivepower2}), we can see that for the tuned line, the reactive power gets reduced. Reactive power is one of the main components that hampers transmitting bulk (active) power in the HVAC system. Therefore, with the tuned HVAC line, the reactive power could expectedly be reduced.

\section{Simulation Results}
Simulation of a tuned HVAC line is performed using the PSCAD/EMTDC \cite{PSCAD}. The simulation is performed in the following way.

\begin{itemize}
\item Transmission line is at 220 kV, supplied by an ideal voltage source (synchronous generator).
\item A capacitive load of 100 MVAr is connected at the receiving side.
\item For the tuned HVAC line, the supply frequency is varied from 50 Hz to 1 kHz.
\item Two sets of experiments are performed with line length of 500 km and 300 km.
\item The active power transmitted and the reactive power absorbed in the line as a function of the supply frequency are noted.
\end{itemize}

\begin{figure}[ht]
\centering
\includegraphics[width=6.7in]{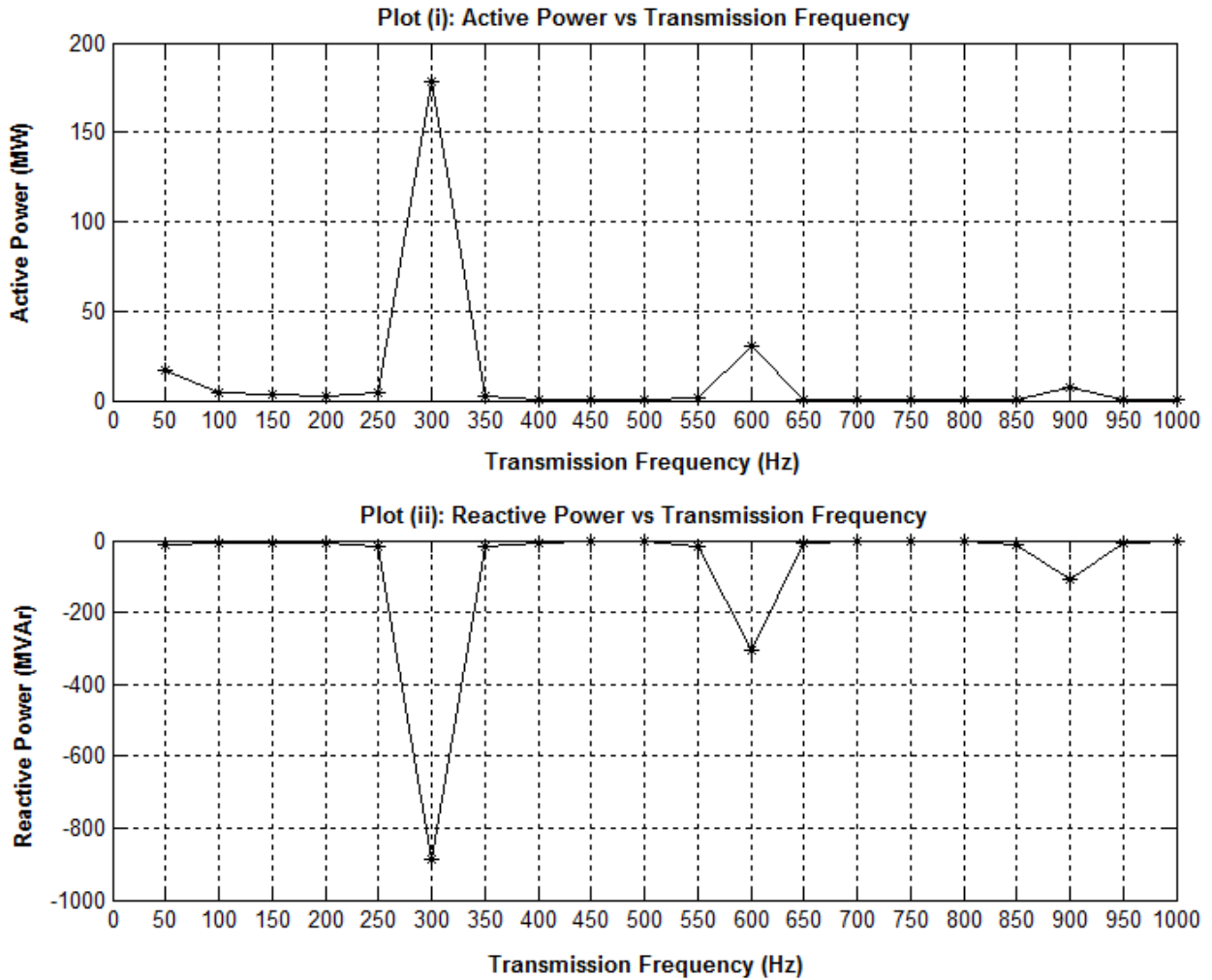}
\caption{Simulation of long ($500$ km) tuned HVAC line: Plot (i): Active power, Plot (ii): Reactive power.}
\label{fig:PQ500}
\end{figure}

For the line length of 500 km, the results are shown in Fig. \ref{fig:PQ500}, plot (i) showing the active power (MW), and plot (ii) showing the reactive power (MVAr) absorbed in the line. As the line length is 500 km, the tuning effect can be noticed at frequencies of 300 Hz, 600 Hz, 900 Hz. From plot (ii), the reactive power gets reduced drastically at the first tuning frequency, while it reduces less in the subsequent tuning frequencies. The active power accordingly increases at the respective tuning frequencies. This supports the argument that the tuned HVAC lines could help to transmit increased amount of power at particular frequency, depending on the length of the line. 

\begin{figure}[ht]
\centering
\includegraphics[width=6.7in]{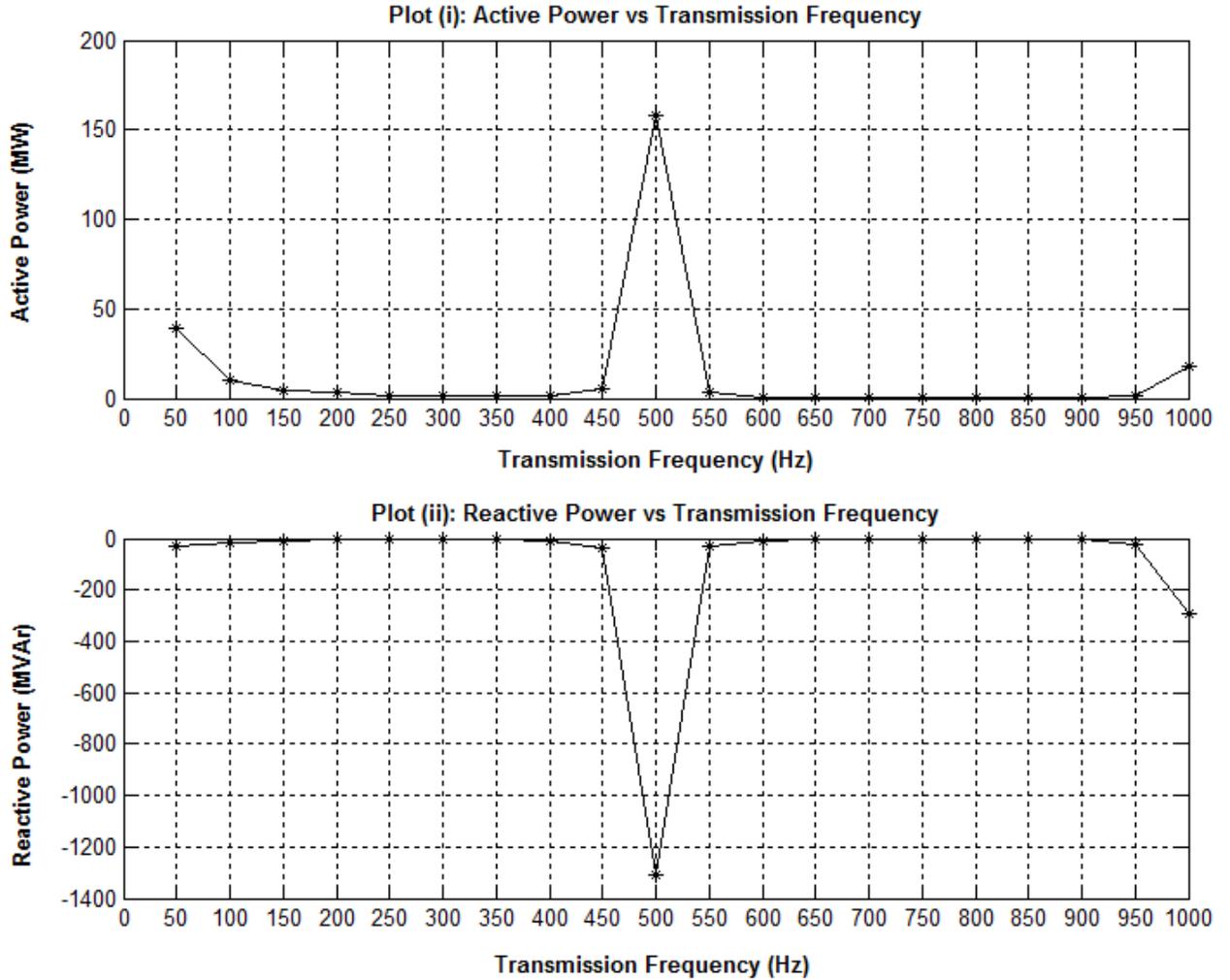}
\caption{Simulation of long ($300$ km) tuned HVAC line: Plot (i): Active power, Plot (ii): Reactive power.}
\label{fig:PQ300}
\end{figure}

For the line length of 300 km, the results are shown in Fig. \ref{fig:PQ300}, plot (i) showing the active power (MW), and plot (ii) showing the reactive power (MVAr) absorbed in the line. As the line length is 300 km, from eq. (\ref{eq:linelengthfreq}), the tuning effect can be noticed at frequencies of 500 Hz, 1 kHz. From plot (ii), the reactive power gets reduced drastically at the first tuning frequency. The active power accordingly increases at the respective tuning frequencies.  

\section{Conclusion}
One of the main objectives of the smart grid initiative is to enable long distance bulk power transmission with reduction in transmission losses. Different technologies are being investigated and implemented, among which HVDC is a major one. In this paper, a concept of tuned HVAC line is presented for long ($>250$ km) transmission line. A tuned line is where the receiving-end voltage and current are numerically equal to the corresponding sending-end values. This paper presents the theoretical analysis of the tuned line, suggesting adaptation of the transmission frequency ($f$) as per the length of the line ($l$), based on the relation $f=\frac{nc}{2l}$ ($c$ is speed of light, $n=1,2,3,\ldots$ is a parameter). This would not require to convert from the AC to the DC system in principle, rather changing the AC frequency would suffice. Therefore, it is essentially an AC system. The simulation of a tuned HVAC line is performed using the PSCAD/EMTDC. Simulation results for the two different line lengths, substantiate the theoretical analysis of reducing the reactive power absorbed in the line, while increasing the active power transmission.

\section*{Acknowledgment}
The work was supported in part by the Start-Up Grant (M4081235.040), Nanyang Technological University, Singapore.

\begin{biography}[{\includegraphics[width=1in,height=1.25in,clip,keepaspectratio]{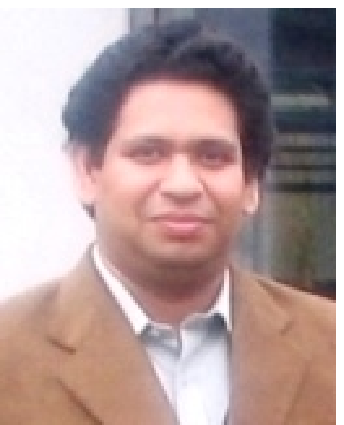}}]{Abhisek Ukil}
received the B.Eng degree in electrical engineering from the Jadavpur Univ., Kolkata, India, in 2000 and the M.Sc. degree in electronic systems and engineering management from the Univ. of Bolton, Bolton, UK in 2004. He received the Ph.D. degree from the Pretoria (Tshwane) University of Technology, Pretoria, South Africa in 2006, working on automated disturbance analysis in power systems.

Since 2013, he is assistant professor in the School of EEE, Nanyang Technological University, Singapore. From 2006--2013, he was Principal Scientist at the Asea Brown Boveri (ABB) Corporate Research Center, Baden-Daettwil, Switzerland. He is author/coauthor of more than 60 refereed papers, a monograph, 2 book chapters, and inventor/co-inventor of 9 patents. His research interests include smart grid, power systems, HVDC, renewable energy \& integration, condition monitoring, signal processing applications.
\end{biography}

\end{document}